# Stochastic Scenario Analysis of Andes Virus Outbreak aboard MV Hondius Cruise Ship, 2026

**Running Title:** Andes Virus Outbreak on MV Hondius

**Authors:** Raj Kumar Subedi[1†]; Hamed Karami[2†]; Kaustubh Wagh[1]; Kenji Mizumoto[3]; Gerardo Chowell[1,4*]

**Affiliations:**

[1]School of Public Health, Georgia State University, Atlanta, GA, USA

[2]Department of Mathematics and Statistics, Georgia State University, Atlanta, GA, USA

[3]Graduate School of Advanced Integrated Studies in Human Survivability, Kyoto University, Yoshida-Nakaadachi-Cho, Sakyo-Ku, Kyoto, Japan

[4]Department of Applied Mathematics, College of Applied Sciences, Kyung Hee University, Korea

[†]Joint first authors

*Corresponding author: Gerardo Chowell, PhD. Email: gchowell@gsu.edu

# Abstract

In April 2026, the MV Hondius cruise ship experienced the first documented cruise ship-associated Andes hantavirus (ANDV) outbreak, with 13 cases (11 confirmed, 2 probable) and 3 deaths in an analytical cohort of 149 passengers and crew. We applied a stochastic epidemic model to evaluate four embarkation scenarios under reproduction numbers anchored to published ANDV estimates. Under the two-beta model, Scenario D (two latent infected persons at embarkation) most often generated outbreaks as large as observed, yielding P(final size ≥ 13) = 11.6% and P(takeoff) = 58.5% at $R_0 = 2.12$. ABC supported a broader multiple-latent embarkation state, although transmissibility and the exact initial state remained weakly identifiable. Under the selected day-35 change point, simulations did not resolve a substantial change in takeoff probability. Scenario modeling of a well-enumerated travel cohort can inform exposure-history assessment, early onboard surveillance, rapid isolation, and post-disembarkation monitoring for long-incubation pathogens in confined travel settings.

# Introduction

Andes hantavirus (ANDV) is the only hantavirus for which the possibility of person-to-person transmission has been documented, typically requiring close prolonged contact *(1, 2)*. The incubation period ranges from 7 to 39 days with a mean of approximately 18 days *(2-4)*, and the case fatality rate approaches 30-40% *(2)*. The 2018-2019 Epuyén outbreak in Argentina demonstrated sustained human-to-human ANDV transmission with estimated reproduction numbers of 0.96 post-intervention and 2.12 pre-intervention *(5)*. These features create particular challenges for travel-associated outbreaks: exposed travelers may remain asymptomatic during transit, develop illness after disembarkation, and disperse internationally before infection is recognized.

Cruise ships are distinct confined travel settings for outbreak investigation, often providing a well-described analytical cohort with documented voyage timing. The Diamond Princess COVID-19 outbreak demonstrated that such settings can yield informative transmission estimates *(6-8)*. However, cruise ship contact patterns are heterogeneous, involving shared cabins, dining groups, excursions, healthcare encounters, and crew-passenger interactions *(9)*. Such settings should not be treated as equivalent to homogeneous shipwide mixing. For ANDV, this distinction is especially important because the incubation period described above spans the duration of the voyage, so infections acquired before boarding may not become apparent until well into or after the sailing. Unlike pathogens with incubation periods of hours to a few days, onset timing alone therefore cannot separate pre-boarding from onboard exposure, and the relevant risk set may include persons with shared pre-boarding environmental exposures as well as persons exposed through close contact during the voyage *(10)*.

As of May 27, 2026, 13 cases of ANDV (11 confirmed and 2 probable) were identified among MV Hondius passengers and crew, constituting the first documented cruise ship-associated ANDV outbreak. Because all 13 cases were epidemiologically linked to a single voyage and ANDV infection is rare, with no expected background occurrence in this population, we refer to this event as an outbreak rather than a cluster, consistent with the terminology subsequently adopted by WHO. Three deaths were recorded (case fatality rate 27.3% among confirmed cases). Current evidence suggests a shared pre-boarding environmental exposure in Argentina with subsequent limited onboard human-to-human transmission *(11, 12)*, though the number of individuals silently incubating at embarkation remained uncertain.

A recent commentary highlighted the broader preparedness implications of the MV Hondius cluster, including the vulnerability of expedition cruise travel to emerging zoonotic threats *(11)*. However, outbreak summaries and commentaries have not quantitatively compared competing embarkation and transmission scenarios using the reported timing information and final outbreak size. We therefore applied a stochastic compartmental model to evaluate competing embarkation scenarios, quantify outbreak probabilities under literature-based reproduction numbers, and assess a counterfactual change in transmission at the time of public health response. Simulation-based inference was used to assess which epidemiologically plausible embarkation states were most compatible with the limited observed data.

# Methods

## Study population and data

We used N = 149 passengers and crew as the selected analytical cohort *(13)*. Case data were compiled from WHO Disease Outbreak News and ECDC rapid risk assessments *(12-15)*. The case count, consistent with official reports, was 13 (11 confirmed, 2 probable); 9 of 13 cases had publicly available symptom-onset dates *(15)*. Four cases detected during post-disembarkation quarantine surveillance lacked public symptom-onset dates and are not displayed in the onset-date epidemic curve (data freeze: May 27, 2026) *(14)*. For ABC calibration, timing was defined as symptom onset when publicly available and report/detection date otherwise, yielding 8 Monday-start weekly calibration bins totaling 13 cases. Because publicly reported voyage and cohort information may not fully capture all embarkations, disembarkations, and post-disembarkation follow-up, N = 149 should be interpreted as a selected analytical denominator rather than proof of a perfectly closed mixing population.

## Stochastic model

We applied a stochastic small-step Poisson tau-leaping $SE_1E_2E_3IR$ model to a constant-population analytical cohort of N = 149. This model was used as an illustrative homogeneous-mixing benchmark rather than a full reconstruction of person-to-person contact networks aboard the ship. The exposed period was divided into n = 3 sequential subcompartments to approximate an Erlang-distributed incubation period *(16)*. The number of subcompartments (n = 3) and progression rate ($\sigma$ = 3/18 per day) were derived from published incubation parameters *(2-4)* (Appendix 1). A two-beta transmission structure separated pre- and post-intervention rates, anchored to published reproduction number estimates *(5)* (Appendix 1). Each scenario was run for 1,000 simulations over $t_{max}$ = 45 days ($\Delta t$ = 0.01 days) *(17)*. The intervention cutoff at day 35 corresponded to ECDC's initial classification of all onboard passengers and crew as high-risk contacts and deployment of WHO medical expertise, which likely initiated behavioral changes including informal distancing and isolation measures prior to formal disembarkation on May 10. Full model specification and population conservation diagnostics are provided in the Appendix. Under ideal continuous noise-free observation, $\beta$ and initial conditions were globally structurally identifiable in the single-beta $SE_1E_2E_3IR$ ODE model *(18, 19)*, assuming N, $\sigma$, and $\gamma$ fixed and S(0), I(0), and R(0) treated as known.

## Scenario definitions

Four scenarios representing alternative embarkation states were evaluated (Table 2). Reproduction numbers were anchored to Martínez et al. 2020 *(5)*: $R_0$ = 0.70 (Scenario A, subcritical) and $R_0$ = 2.12 (Scenarios B-D, pre-intervention). A sensitivity sweep across $R_0 \in \{0.96, 1.19, 2.12\}$ was performed for Scenarios B, C, and D in the single-beta model (Appendix Table 6); in the two-beta model, where $R_1$ = 0.96 is fixed from day 35, the sweep covered $R_0 \in \{1.19, 2.12\}$ (Figure 3). These values are effective reproduction numbers for the settings in which they were estimated or assumed, and $R_0$ is used here to denote the scenario- or period-specific value rather than a pathogen-level basic reproduction number. Final outbreak size included both infected individuals present at embarkation and incident infections generated during simulation. Structural identifiability does not imply practical identifiability in the presence of observation noise and small case counts (see Results).

## Simulation-based inference

ABC with rejection sampling *(20-22)* was applied to jointly estimate $R_0$ and initial conditions $E_1(0)$ and $E_3(0)$ from weekly calibration-date incidence. ABC simulations were implemented using the Epydemix toolbox *(21)*, which uses a discrete-time stochastic multinomial compartmental simulator. In the ABC analysis, transmission was parameterized using a single constant $\beta = R_0 \gamma$ over the calibration window; the day-35 change point used in the separate scenario analysis was not applied. Simulated weekly incidence was defined as the number of $E_3 \rightarrow I$ transitions (onset incidence), matching the timing variable used for calibration; $S \rightarrow E_1$ infection incidence was not used in the ABC distance. The distance metric was weekly incidence RMSE; the top 1% of 100,000 simulations were retained as the approximate posterior *(23)*. The observed ABC vector was [1, 0, 2, 5, 1, 2, 1, 1] across 8 Monday-start weekly bins from April 6 through May 25, 2026, totaling 13 cases. Calibration dates were defined as symptom-onset dates when publicly available and report/detection dates for the four cases without public symptom-onset dates. The primary prior was $R_0 \sim U(0.5, 3.0)$, with sensitivity analysis under $R_0 \sim U(0.5, 4.0)$; $E_1(0) \sim$ Discrete{0,1} and $E_3(0) \sim$ Discrete{0,1,2}, with I(0) fixed at 0. Because the zero-seed state cannot generate an outbreak unless spontaneous importation is separately modeled, posterior support over initial conditions was interpreted relative to viable seeded states. Because I(0) was fixed at 0, this ABC specification evaluated latent-infection embarkation states rather than scenarios with an infectious individual already present at embarkation.

## Statistical analysis

Primary descriptive outcomes were P(final size ≥ 13) and P(takeoff, defined as final size ≥ 5) at day 45. The simulation window of 45 days captured the full observed onset distribution (maximum onset day 39) with a 6-day buffer. P(final size ≥ 13) is a threshold probability rather than the likelihood of observing exactly 13 cases. Because simulations stop at day 45 rather than running to absorption, final size should be interpreted as cumulative infections by day 45. Temporal agreement with the observed calibration-date pattern was assessed using weekly-incidence RMSE against the calibration-date vector described above. All scenario-model analyses were conducted in R version 4.4.2 *(24)*.

# Results

## Descriptive epidemiology

Between April 6 and May 25, 2026 (data freeze May 27, 2026), 13 ANDV cases were identified among MV Hondius passengers and crew (Table 1, Figure 1). Eleven cases were passengers and two were crew members. Passenger- and crew-specific denominators were not publicly reported, so proportions in Table 1 are expressed relative to the 149-person cohort rather than as group-specific attack rates. Three deaths occurred (CFR 27.3% among confirmed cases). Symptom onsets in the 9 dated cases ranged from day 5 to day 39 since embarkation (mean 25.4 days, median 27.0 days). The later onsets are compatible with infection acquired onboard. The earliest onset, on day 5, falls below the shortest ANDV incubation period yet reported *(3-5)*, so that case was most plausibly infected before embarkation. The scenario analysis that follows addresses how many such infections would be needed to reproduce an outbreak of the observed size.

## Stochastic scenario results

Under the two-beta model, Scenario D ($E_1(0) = 1$, $E_3(0) = 1$, $I(0) = 0$ at embarkation; $R_0 = 2.12$, $R_1 = 0.96$) yielded the highest probability of generating an outbreak at least as large as the observed outbreak size: P(final size ≥ 13) = 11.6% and P(takeoff) = 58.5% (Figure 2, Appendix Tables 4, 5). Scenarios B and C produced intermediate threshold probabilities (P(final size ≥ 13) = 7.6% and 4.2%, respectively). Scenario A (subcritical) produced P(final size ≥ 13) = 0% as expected. Across all $R_0$ values in the sensitivity sweep, Scenario D consistently yielded the highest probability of crossing the final-size threshold (Figure 3). Unconditioned and conditioned trajectories and the corresponding epidemic size distributions are shown in Appendix Figures 5-7. At $R_0 = 0.96$ (post-intervention), all scenarios approached zero (Appendix Table 6), consistent with the potential effectiveness of isolation measures had they been implemented earlier.

## Intervention impact

Under the selected day-35 change point and parameter assumptions, simulations did not resolve a substantial difference in epidemic takeoff probability across scenarios, with differences of ≤1.1 percentage points between single-beta and two-beta models (Appendix Figures 8 and 9; Appendix Table 7). With 1,000 simulations per condition, differences of this magnitude are within Monte Carlo variation. This finding is consistent with the long mean incubation period: by day 35, much of the subsequent case burden may already have been embedded in latent infections. In this counterfactual, late symptom-triggered interventions had limited detectable effect once infections were already incubating.

### Simulation-based inference

ABC jointly estimated $R_0$ and initial conditions from the 8-bin weekly calibration-date incidence vector totaling 13 cases. Across both prior specifications (upper bounds 3.0 and 4.0; 100,000 simulations; 1% acceptance; Appendix Table 8), the $R_0$ posterior consistently approached the prior upper boundary, indicating weak identifiability from the available data, as expected given the small outbreak size. In contrast, the joint posterior over initial conditions supported a broader multiple-latent embarkation state. Under the original unconditional prior, the most supported pair was $E_1(0) = 1$ and $E_3(0) = 2$ (33.8% under U(0.5, 3.0); Figure 4; Appendix Table 9, Appendix Figures 10, 11), corresponding to three latent infections at embarkation. Under the viable-seed conditional prior, this joint state and the marginal posterior support for $E_3(0) = 2$ increased relative to prior expectation, whereas the marginal support for $E_1(0) = 1$ was similar to its viable-seed prior probability. ABC therefore provides complementary support for multiple-latent infections at embarkation, but the result should not be interpreted as a definitive reconstruction because timing was uncertain for cases without public symptom-onset dates and I(0) was fixed at 0 (Appendix Figure 12).

## Discussion

We conducted a stochastic scenario analysis of the MV Hondius ANDV outbreak using public-source outbreak data. The day-5 onset points to at least one infection before embarkation on descriptive grounds, and the scenario analysis addresses how many such infections are needed to reproduce the observed outbreak. The observed outbreak pattern was most dynamically compatible with multiple-latent infections present at embarkation under the prespecified homogeneous-mixing benchmark model. ABC provided complementary support for a broader multiple-latent embarkation state, with strongest support for E1(0) = 1 and E3(0) = 2 across both prior specifications.

The selected day-35 intervention counterfactual did not resolve a substantial change in P(takeoff), with differences of ≤1.1 percentage points under the simulation settings used here. The finding is compatible with a fundamental challenge in ANDV outbreak management: the long mean incubation period means that by the time cases are identified and interventions implemented, a substantial portion of the subsequent case burden may already be embedded in silently incubating infections. This interpretation supports the importance of pre-embarkation exposure history assessment and early onboard surveillance, which are core elements of established international guidance for communicable disease surveillance and control on passenger ships *(25, 26)*, because symptom-based interventions implemented after outbreak recognition may have limited detectable impact when infections are already incubating. For public health response, this underscores the importance of post-disembarkation monitoring, contact tracing across international borders, and rapid alert to receiving countries when travelers from an ANDV exposure event disperse *(27)*.

Our two-beta framework quantitatively separates pre- and post-intervention transmission dynamics, allowing explicit assessment of intervention timing under a simplified counterfactual model. This framework is applicable to future closed-population or selected-cohort outbreak investigations with documented intervention dates, complementing earlier modeling of countermeasure timing in shipboard outbreaks *(28)*. Population conservation diagnostics indicate that boundary corrections were rare and small in these runs (≤0.0002% of time steps, maximum correction of 1 individual; Appendix Tables 2, 3; Appendix Figures 1-4).

The ABC analysis provided a useful complementary check on the scenario results. The analysis supported the idea that more than one silently infected person may have boarded the ship, with strongest support for one early-latent and two late-latent infections at embarkation. Conditioning the prior on viable seeded states, which removes the non-generative zero-seed state from the comparison, strengthened support for two late-stage latent infections; transmission intensity was less well resolved, as expected given the small outbreak size.

We acknowledge some limitations. First, the model assumes homogeneous mixing within a selected analytical cohort, which may not fully capture heterogeneous contact patterns across cabins, dining groups, and excursion subgroups *(10, 11)*. Second, the constant-population approximation may not fully reflect embarkations, disembarkations, and post-disembarkation infection progression. Third, four cases had no reported onset dates; for ABC calibration, these cases were represented using report/detection dates, which may differ from true symptom onset and could affect timing-based inference. Fourth, $R_0$ values are anchored to a community outbreak in Argentina *(5)* and may not fully reflect transmission intensity in the confined cruise ship setting. Fifth, the small outbreak size limits statistical precision across all analyses. Sixth, because scenarios with more initially infected individuals are more likely to reach larger final sizes by construction, scenario comparisons should be interpreted as plausibility screening rather than formal proof of the true transmission pathway; relatedly, the day-35 two-beta comparison is a model-based counterfactual under fixed literature-based parameters rather than an evaluation of the actual public health response, and transmissibility was only weakly identifiable from the 13-case series. Finally, our data freeze was May 27, 2026. Subsequent public updates reported the same total of 13 cases, with one case reclassified from probable to confirmed, and WHO declared the outbreak over on 2 July 2026 *(29)*. Because the final outbreak size was unchanged, the analyses reported here are unaffected.

The MV Hondius outbreak illustrates both the opportunities and challenges of selected-cohort outbreak analysis in travel settings. The documented voyage timing and available denominator support scenario-based modeling; the small case count, missing onset dates, and possible changes in the onboard population limit transmission parameter identifiability. Passenger ships have a long documented record of communicable disease transmission, most extensively for gastrointestinal, respiratory, and vaccine-preventable infections *(30, 31)*, but expedition vessels operating in remote regions raise distinct preparedness questions for emerging zoonoses *(11, 25)*. Future investigations should prioritize individual-level contact data, cabin and excursion records, and genomic sequencing to discriminate between shared ecological exposure and secondary transmission pathways *(11)*. The key travel and outbreak medicine implication is not the exact number of infected persons at embarkation, but the need for early recognition, exposure-history assessment, rapid isolation, appropriate protection for crew and medical staff during close contact, and post-disembarkation monitoring when travelers from endemic regions disperse internationally during a long incubation period.

## Conclusions

The MV Hondius ANDV outbreak was compatible with multiple incubating infections at embarkation, though the analysis addresses how many latent infections were plausibly present, not how they arose. In the counterfactual model, a day-35 transmission reduction did not resolve a substantial change in epidemic takeoff probability, underscoring the importance of exposure-history assessment, early onboard surveillance, rapid isolation of symptomatic cases, and post-disembarkation monitoring for ANDV management in selected-cohort settings. Applied prospectively, this form of scenario screening needs the case count, cohort size, and voyage timing, together with an assumed $R_0$ range and incubation/infectious-period parameters, and can indicate early whether an observed outbreak is plausible from onboard transmission alone or instead requires infections already present at embarkation. Simulation-based inference provided complementary support for a broader multiple-latent embarkation state, while identifying $R_0$ as weakly identifiable.

## Author Contributions

G.C. and K.M. conceptualized the study and provided the foundational modeling framework. R.K.S. adapted and executed the R implementation of the stochastic epidemic models, incorporating structural modifications and corrections. H.K. developed and performed the Approximate Bayesian Computation (ABC) analysis. R.K.S. and H.K. drafted the first version of the manuscript. K.W. and K.M. contributed to the interpretation of the mathematical simulations and participated in the critical revision of the manuscript. All authors reviewed, edited, and approved the final version of the manuscript.

## Acknowledgements

We thank WHO, ECDC, and national health authorities for transparent outbreak reporting, and the Virological.org community for rapid genomic characterization.

## Funding

None declared.

## Conflict of interest

The authors declare no competing interests.

## Ethics statement

This study used publicly available, de-identified aggregate outbreak data; ethics review was not required.

## Data availability

Case data *(15)* were compiled from publicly available WHO and ECDC outbreak reports. Analysis approaches and model specifications are described in the Appendix and available in GitHub repository *(14)*.

## Tables

**Table 1.** Descriptive summary of the MV Hondius Andes virus outbreak, 2026, Data freeze May 27, 2026.

| **Parameter** | **Value** |
|---|---|
| Total cases | 13 |
| Confirmed | 11 |
| Probable | 2 |
| Deaths | 3 |
| CFR - confirmed (%) | 27.3 |
| Cases with onset dates | 9 |
| Cases among passengers, n (% of 149-person cohort) | 11 (7.4) |
| Cases among crew, n (% of 149-person cohort) | 2 (1.3) |
| Mean days since embarkation to onset (n=9) | 25.4 |
| Median days since embarkation to onset (n=9) | 27.0 |

Table 1 source: This study as of May 27, 2026 data. All deaths were among confirmed cases.

**Table 2.** Prespecified stochastic scenario definitions, MV Hondius Andes virus outbreak, 2026

| Scenario | Description | $E_1$(0) | $E_3$(0) | I(0) | $R_0$ |
|---|---|---|---|---|---|
| A | Late-stage latent seed, subcritical transmission | 0 | 1 | 0 | 0.70 |
| B | Infectious seed at embarkation, no latent co-exposure | 0 | 0 | 1 | 2.12 |
| C | Single late-stage latent seed, supercritical transmission | 0 | 1 | 0 | 2.12 |
| D | Two latent seeds, early- and late-stage | 1 | 1 | 0 | 2.12 |

Table 2 source: This study as of May 27, 2026 data. $R_1$ = 0.96 was applied from day 35 (May 6, 2026) in all two-beta scenarios, with $\beta_1 = R_1\gamma$. Final outbreak size included both infected individuals present at embarkation and incident infections generated during simulation.

Scenarios were treated as epidemiologically plausible hypotheses, not inferred transmission states. In Scenario A ($R_0 = 0.70 < R_1 = 0.96$) the day-35 change point does not represent an intervention; because almost no simulations retain an infectious individual by day 35, single-beta and two-beta results for A are identical (Appendix Table 7), and the two-beta specification should be interpreted as a counterfactual intervention only for Scenarios B-D.

# Figures

**Figure 1.** Epidemic curve by reported symptom-onset date and case status, MV Hondius Andes virus outbreak, 2026. Bars show the 9 confirmed/probable cases with publicly reported onset dates; four additional confirmed/probable cases lacked reported onset dates and are not displayed. Vertical dashed line indicates cruise embarkation from Ushuaia on April 1, 2026. Source: WHO/ECDC public updates, data as of May 27, 2026.

**Figure 2.** Stochastic Poisson $SE_1E_2E_3IR$ trajectories conditioned on epidemic takeoff (final size ≥ 5) under the two-beta model. Solid lines show median simulated cumulative infections; darker bands show the interquartile range and lighter bands show the 2.5th-97.5th percentile simulation envelope. Dashed line shows the observed cumulative onset curve among cases with known onset dates (n = 9). Dotted vertical line indicates the intervention cutoff, May 6, 2026 (day 35). Scenario A is not shown because too few simulations reached the takeoff threshold.

**Figure 3.** $R_0$ sensitivity: P(final size ≥ 13) across $R_0 \in \{1.19, 2.12\}$ for Scenarios B, C, and D, two-beta model. $R_0$ values from Martínez et al. 2020. Post-intervention $\beta_1$ was set as $R_1 \times \gamma$, with $R_1 = 0.96$ from day 35 (May 6, 2026); $R_0 = 0.96$ is therefore not swept in the two-beta model and is reported for the single-beta model in Appendix Table 6.

**Figure 4.** Approximate Bayesian computation posterior support over initial conditions $E_1(0) \times E_3(0)$ from ABC simulation-based inference. ABC used 100,000 simulations; the top 1% by weekly incidence RMSE were retained as the approximate posterior. The observed vector was [1, 0, 2, 5, 1, 2, 1, 1] across 8 Monday-start weekly calibration bins totaling 13 cases, with calibration dates defined as symptom onset when publicly available and report/detection date otherwise. The prior was $R_0 \sim U(0.5, 3.0)$. The most supported pair was $E_1(0) = 1$ and $E_3(0) = 2$,

with posterior probability 33.8%, corresponding to a multiple-latent embarkation state. The U(0.5, 4.0) sensitivity gave the same posterior mode and is reported in the Appendix.

**Author biographical sketch**: Raj Kumar Subedi is a doctoral candidate and a 2CI Fellow in the Department of Population Health Sciences, School of Public Health, Georgia State University, Atlanta, Georgia, USA. His research interests include infectious disease modeling, epidemic forecasting, and excess mortality analysis.

# Appendix

# Stochastic Scenario Analysis of Andes Virus Outbreak aboard MV Hondius Cruise Ship, 2026

## Methods

### Appendix 1. Full $SE_1E_2E_3IR$ model specification

The exposed period was divided into n = 3 sequential subcompartments to approximate an Erlang-distributed incubation period *(1)*. The number of subcompartments (n = 3) and progression rate (σ = 3/18 per day) were derived from published incubation parameters.

$$n = \frac{\mu^2}{\mathrm{SD}^2} \approx 3$$

where μ = 18 days *(2-4)* and SD = 10 days, giving an exposed-stage progression rate of σ = 3/18 per day, so that the mean exposed duration is n/σ = 18 days. A two-beta transmission structure accounted for the public health intervention at day 35 (May 6, 2026):

$$\beta(t) = \begin{cases} \beta_0 & t < 35 \\ \beta_1 & t \geq 35 \end{cases}$$

where $\beta_0 = R_0\gamma$, $\beta_1 = R_1\gamma$, $R_1 = 0.96$ (post-intervention)*(5)*, and γ = 1/8 per day *(3)*. Each scenario was run for 1,000 simulations over $t_{max}$ = 45 days (Δt = 0.01 days *(6)*).

At each time step $\Delta t = 0.01$ days, five Poisson-distributed random variables give the number of transition events occurring during the interval [t, t+Δt):

$$e_{v1} \sim Poisson\left(\beta(t) \cdot \frac{S \cdot I}{N} \cdot \Delta t\right) S \rightarrow E_1$$

$$e_{v2} \sim Poisson(\sigma \cdot E_1 \cdot \Delta t) E_1 \rightarrow E_2$$

$$e_{v3} \sim Poisson(\sigma \cdot E_2 \cdot \Delta t) E_2 \rightarrow E_3$$

$$e_{v4} \sim Poisson(\sigma \cdot E_3 \cdot \Delta t) E_3 \rightarrow I$$

$$e_{v5} \sim Poisson(\gamma \cdot I \cdot \Delta t) I \rightarrow R$$

Compartment sizes updated with boundary correction:

$$S(t + \Delta t) = max(S(t) - e_{v1}, 0)$$

and analogously for $E_1, E_2, E_3, I, R$. Latent E compartments represented incubating, noninfectious persons; only I contributed to transmission.

**Appendix Table 1.** Full model parameter definitions, values, and sources.

| **Parameter** | **Value** | **Source** |
|---|---|---|
| Total population (N) | 149 | ECDC *(7, 8)* |
| Mean incubation used in model (μ) | 18 days | Vial et al. 2006 *(2)* |
| E subcompartments (n) | 3 | Linear chain trick *(1)* |

| **Parameter** | **Value** | **Source** |
|---|---|---|
| E-stage progression rate (σ) | 3/18 $day^{-1}$ | Linear chain trick *(1)* |
| Mean infectious period (1/γ) | 8 days | Vial et al. 2023 *(3)*, Thaledi et al. 2026 *(9)* |
| R post-intervention ($R_1$) | 0.96 | Martínez et al. 2020 *(5)* |
| R overall outbreak | 1.19 | Martínez et al. 2020 *(5)* |
| R pre-intervention ($R_0$) | 2.12 | Martínez et al. 2020 *(5)* |
| Intervention cutoff (day) | 35 | May 6, 2026 |
| Time step (Δt) | 0.01 days | Gillespie 2001 *(6)* |
| Simulation duration (t_max) | 45 days | This study |
| Simulations per scenario | 1,000 | This study |
| Epidemic takeoff threshold | ≥5 cases | This study |

## Appendix 2. Population conservation diagnostic

**Appendix Table 2.** Population conservation diagnostic across 4,500,000 time steps per scenario, single-beta model.

| Scenario | Steps fired | Fire rate (%) | Sims affected | Max correction | Max N deviation |
|---|---|---|---|---|---|
| A | 4 | 0.0001 | 4/1000 | 1 | 1 |
| B | 4 | 0.0001 | 4/1000 | 1 | 1 |
| C | 6 | 0.0001 | 6/1000 | 1 | 1 |
| D | 3 | 0.0001 | 3/1000 | 1 | 1 |

**Appendix Table 3.** Population conservation diagnostic across 4,500,000 time steps per scenario, two-beta model.

| Scenario | Steps fired | Fire rate (%) | Sims affected | Max correction | Max N deviation |
|---|---|---|---|---|---|
| A | 4 | 0.0001 | 4/1000 | 1 | 1 |
| B | 4 | 0.0001 | 4/1000 | 1 | 1 |
| C | 7 | 0.0002 | 7/1000 | 1 | 1 |
| D | 5 | 0.0001 | 5/1000 | 1 | 1 |

*Median firing rate and median N deviation were zero across all simulations and both model variants. Max correction and max N deviation were 1 individual in all cases. These diagnostics*

*indicate that boundary corrections were rare and small in these runs; they are not a proof of exact population conservation by the update algorithm*.

**Appendix Figure 1.** max() boundary correction firing rate distribution by scenario, single-beta model. Each histogram shows the distribution across 1,000 simulations of the percentage of time steps where at least one compartment required the max(x,0) boundary correction. MV Hondius ANDV Outbreak, 2026 | N=149 | dt=0.01 days.

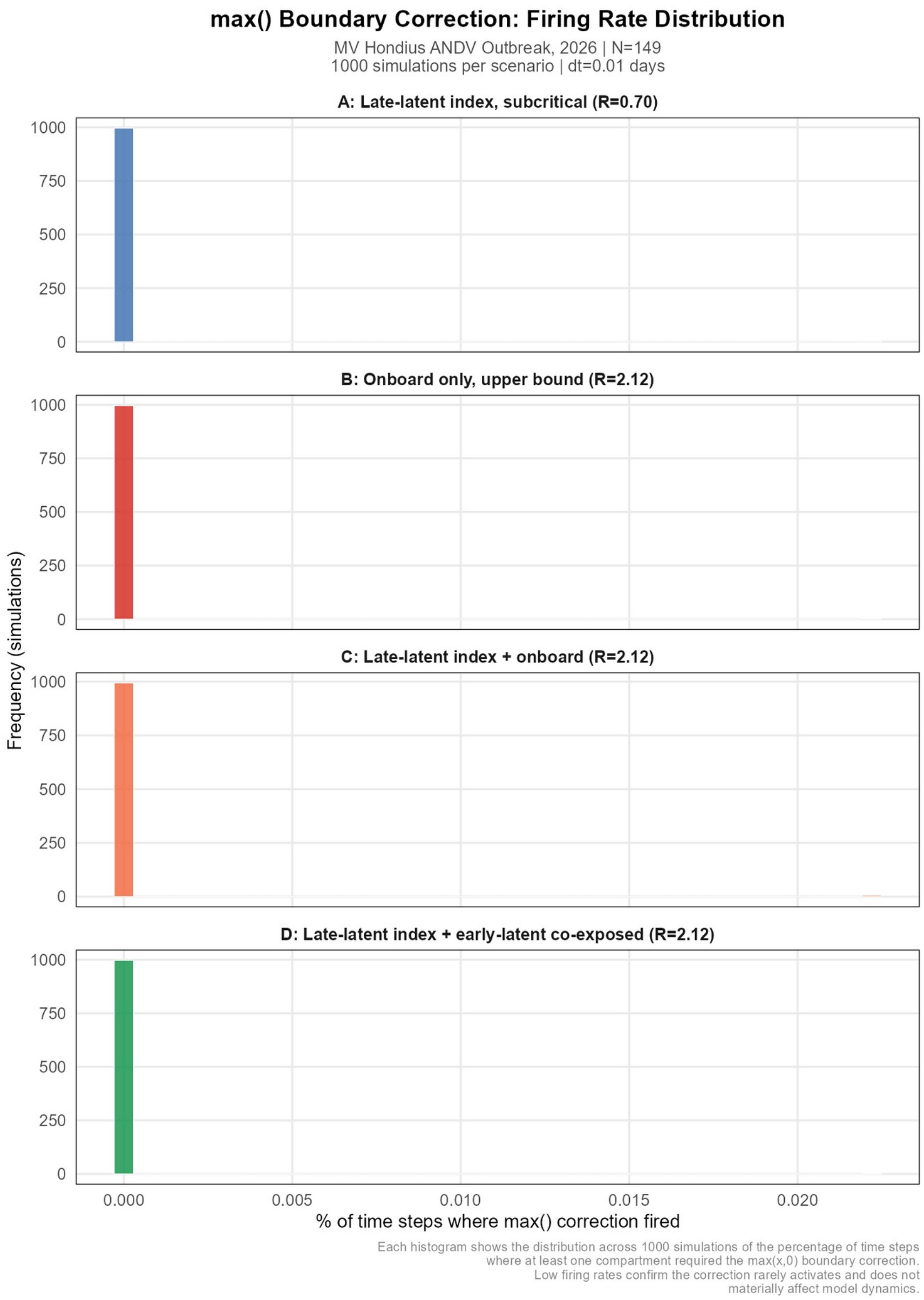

**Appendix Figure 2.** max() boundary correction firing rate distribution by scenario, two-beta model. As Appendix Figure 1 but with $\beta_1 = R_1 \cdot \gamma$ fixed from day 35.

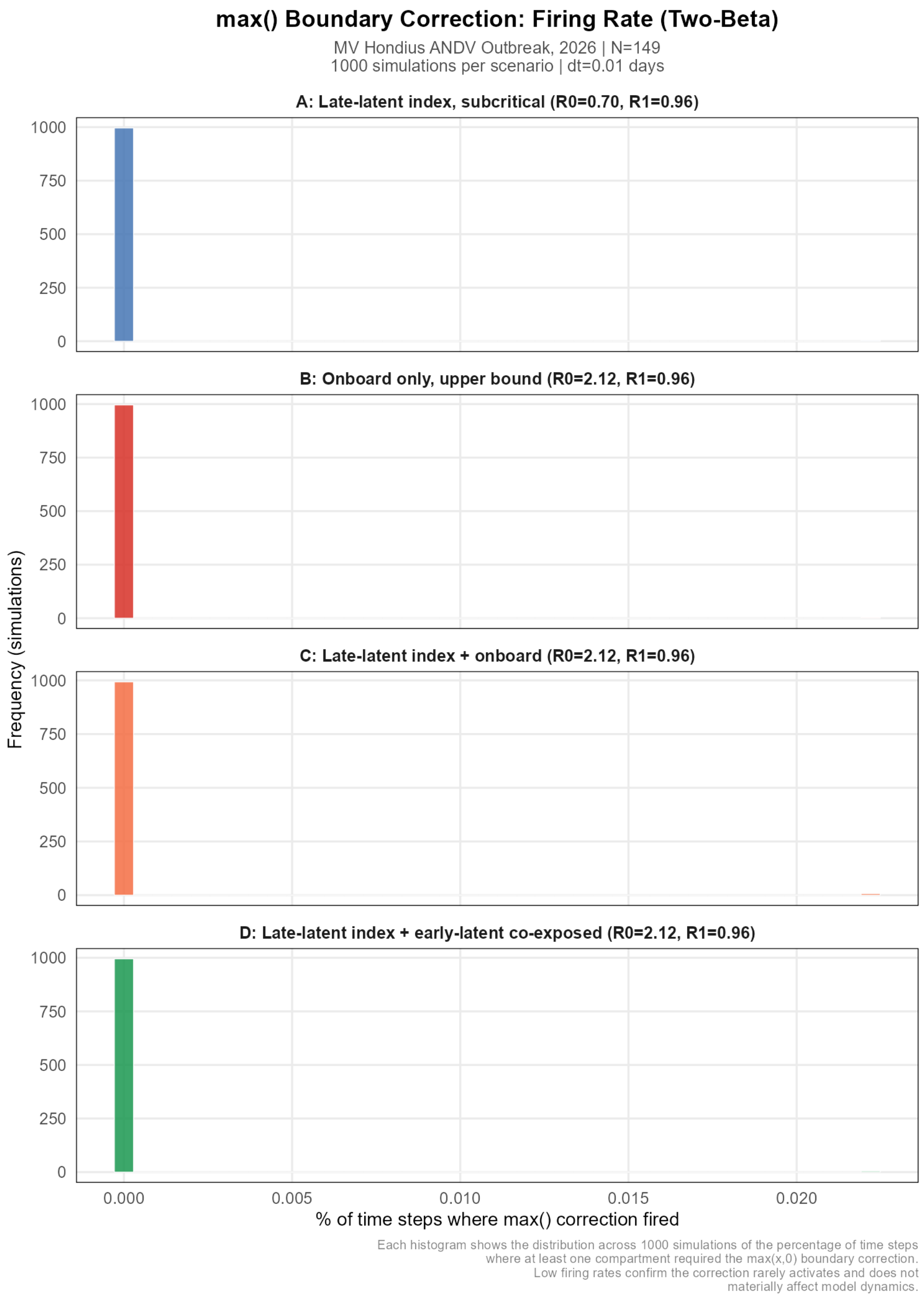

**Appendix Figure 3.** Population conservation: maximum N deviation per simulation by scenario, single-beta model. Each histogram shows the largest single-step deviation from N=149 across all time steps within each simulation. Deviations near zero indicate that boundary corrections were rare and small in these simulations.

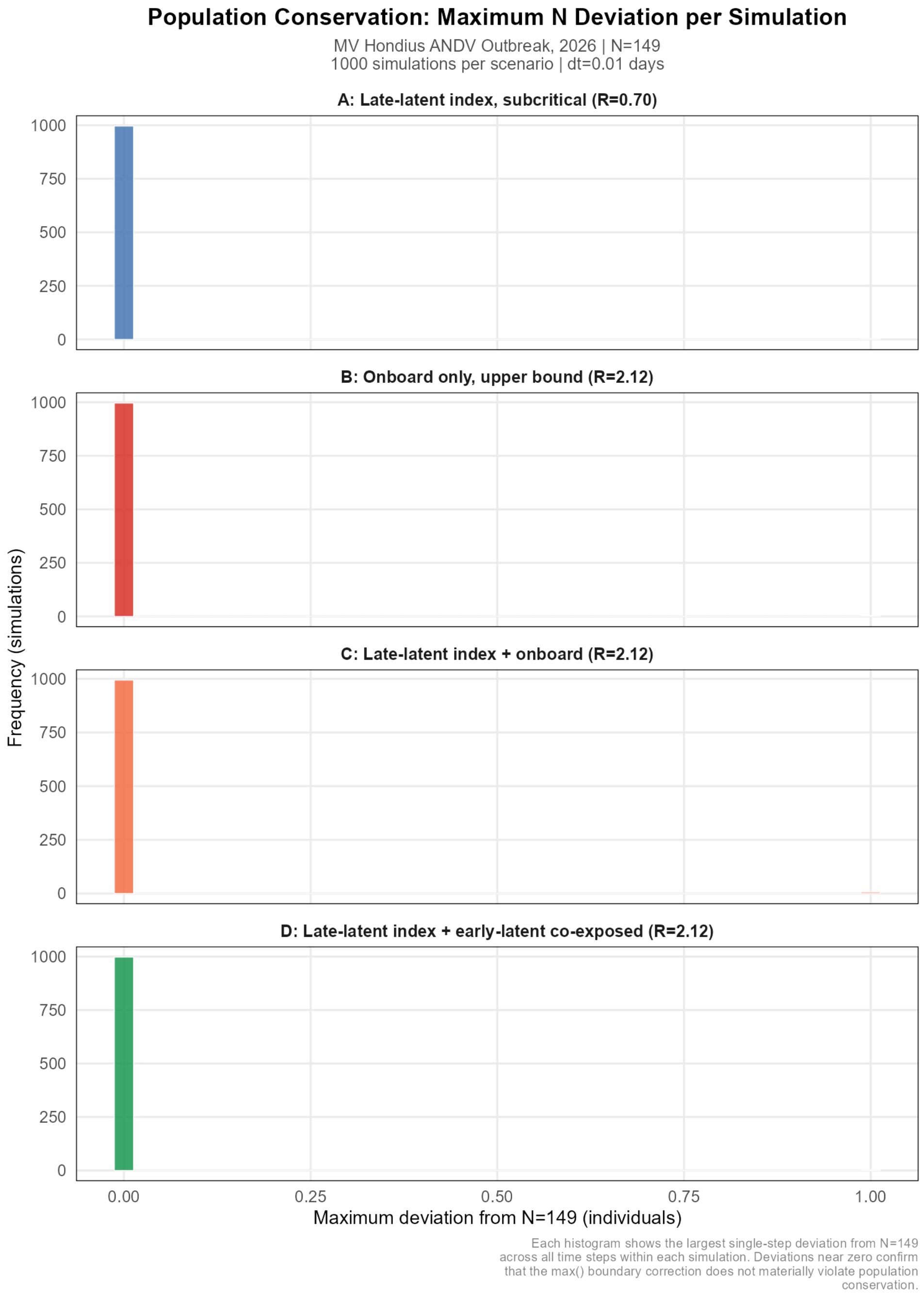

**Appendix Figure 4.** Population conservation: maximum N deviation per simulation by scenario, two-beta model. As Appendix Figure 3 but with two-beta parameterization.

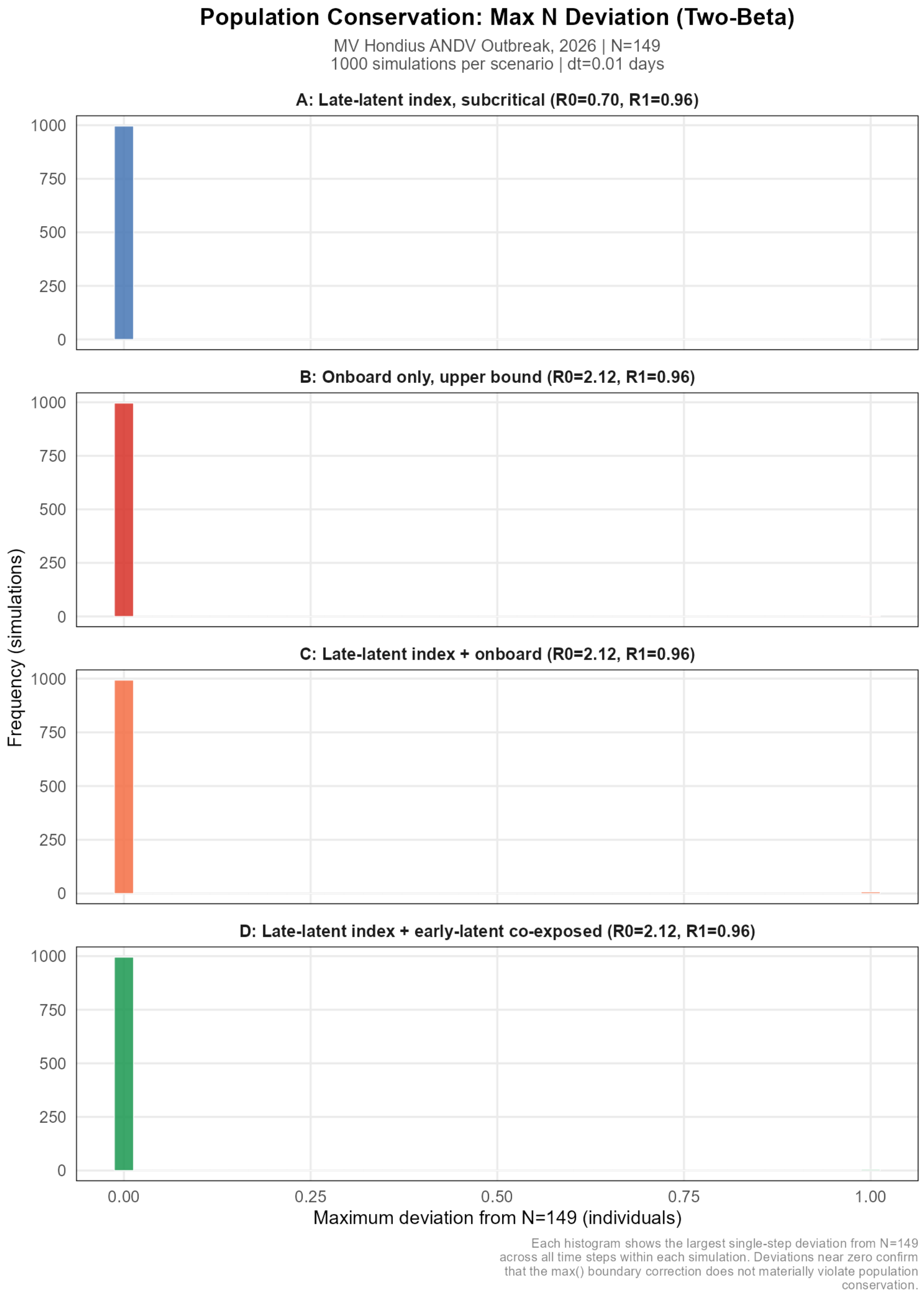

## Appendix 3. ABC implementation details

ABC was implemented in Epydemix v1.2.1 (10) using a stochastic SE1E2E3IR simulator with a single constant transmission parameter, β = R0γ, over the calibration window. The day-35 change point used in the separate scenario analysis was not applied in the ABC analysis. Simulated weekly incidence was defined as the number of E3 → I transitions (onset incidence), matching the timing variable used for calibration; S → E1 infection incidence was not used in the ABC distance. The distance metric was RMSE between observed and simulated weekly incidence across 8 Monday-start weekly calibration bins spanning April 6, 2026 through May 25, 2026, using the observed vector [1, 0, 2, 5, 1, 2, 1, 1] (total = 13), with calibration dates defined as symptom onset when publicly available and report/detection date otherwise. A composite distance metric additionally conditioning on observed final size and peak was considered but excluded because it introduces circularity: acceptance would partly depend on the outcome being predicted. The top 1% of 100,000 simulations were retained as the approximate posterior (11-13). Two R0 prior upper bounds were evaluated: U(0.5, 3.0) and U(0.5, 4.0).

## Appendix 4. Structural Identifiability Analysis

Structural identifiability of the single-beta $SE_1E_2E_3IR$ model was assessed using StructuralIdentifiability.jl in Julia *(14-16)*. Population size ($N = 149$), progression rate ($\sigma = 3/18$ day$^{-1}$), and recovery rate ($\gamma = 1/8$ day$^{-1}$) were treated as fixed. The transmission rate $\beta = R_0 \cdot \gamma$ and the E-stage initial conditions were treated as unknown with S(0), I(0), R(0) treated as known. The observed output was the incidence rate $y(t) = \sigma E_3(t)$.

using StructuralIdentifiability

```
ode_single_beta = @ODEmodel(

    S'(t)  = -R0 * (1//8) * S(t) * I(t) / 149,

    E1'(t) = R0 * (1//8) * S(t) * I(t) / 149 - (1//6) * E1(t), #1//6 is σ = 3/18

    E2'(t) = (1//6) * E1(t) - (1//6) * E2(t),

    E3'(t) = (1//6) * E2(t) - (1//6) * E3(t),

    I'(t)  = (1//6) * E3(t) - (1//8) * I(t),

    R'(t)  = (1//8) * I(t),

    y1(t)  = (1//6) * E3(t)

)

assess_identifiability(ode_single_beta, known_ic = [S, I, R])
```

**Result:** $R_0$ and the unknown initial conditions E1(0), E2(0), and E3(0) returned globally identifiable, conditional on S(0), I(0), and R(0) being treated as known, confirming that under ideal continuous noise-free observation of the incidence curve, these parameters are uniquely recoverable. This is a structural result under idealized continuous observation and does not imply practical identifiability: with 8 weekly counts totaling 13 cases, $R_0$ was only weakly identifiable in the ABC analysis, as expected for an outbreak of this size.

# Appendix Results - Stochastic Model

**Appendix Figure 5.** All simulations (unconditioned), single-beta. MV Hondius ANDV Outbreak, 2026 | N=149 | 1000 simulations per scenario | n=13 confirmed and probable cases (ECDC, data as of May 27, 2026).

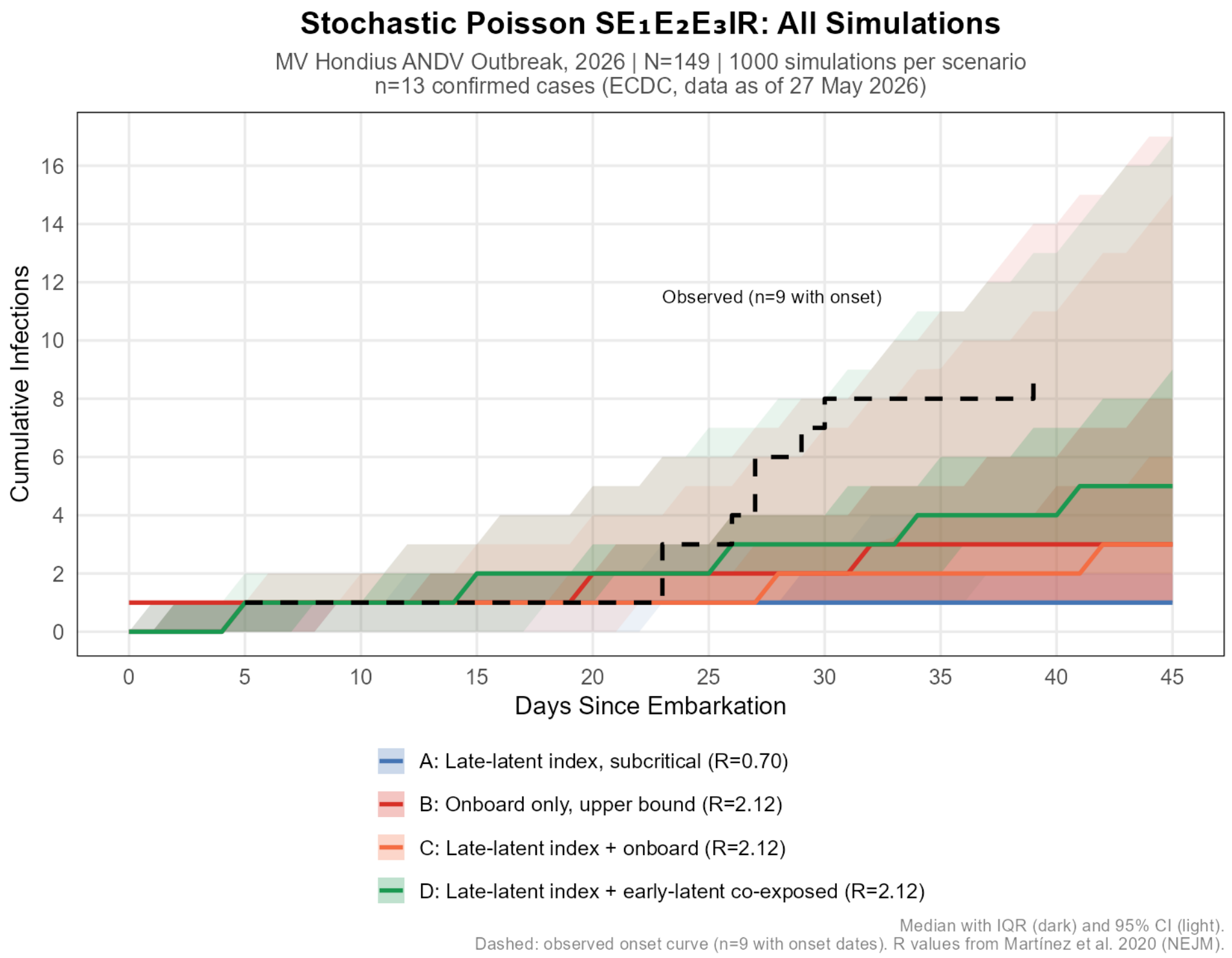

**Appendix Figure 6.** Conditioned trajectories (final size ≥5), single-beta. MV Hondius ANDV Outbreak, 2026 | Excluded: A (insufficient takeoff simulations).

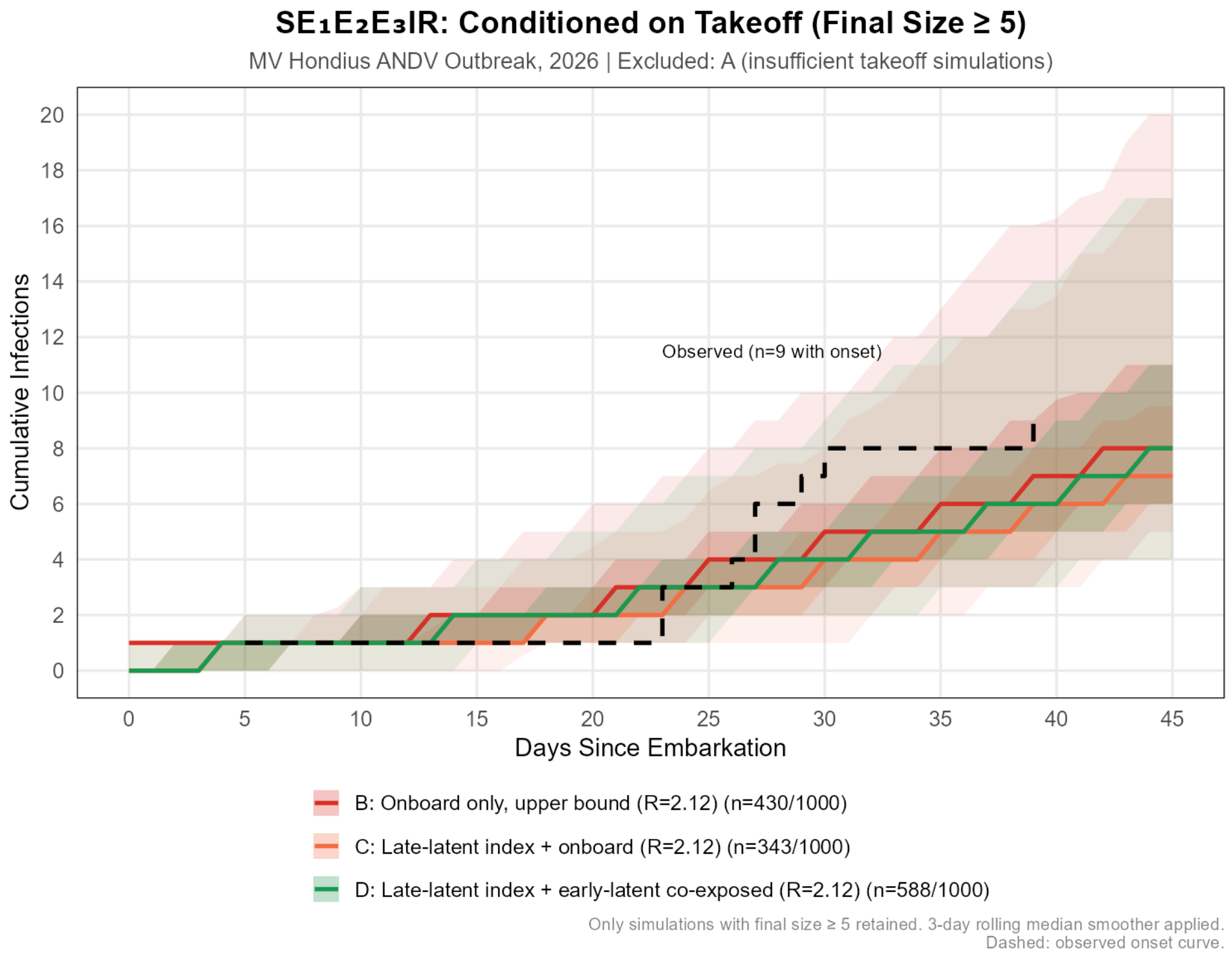

**Appendix Figure 7.** Epidemic size distributions by scenario, single-beta. Day 45 | 1000 simulations per scenario | Observed n=13 confirmed and probable cases (ECDC, data as of May 27, 2026). Dashed: observed n=13.

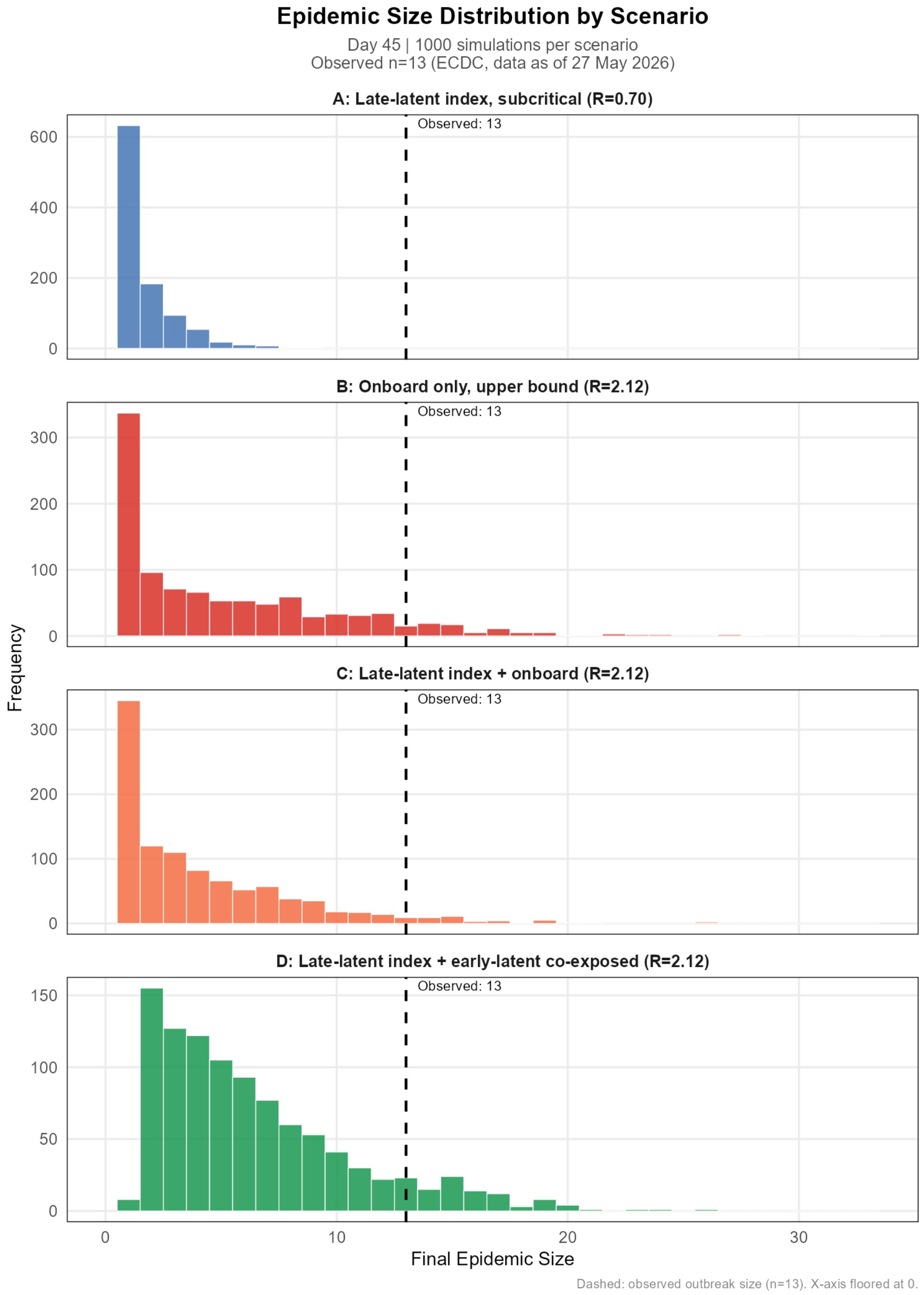

**Appendix Table 4.** Base scenario results, single-beta (n_obs=13, N=149).

| Scenario | Median [IQR] | P(≥13) | P(takeoff) | P(no takeoff) |
|---|---|---|---|---|
| A | 1 [1-2] | 0.000 | 0.037 | 0.963 |
| B | 3 [1-8] | 0.090 | 0.430 | 0.570 |
| C | 3 [1-6] | 0.046 | 0.343 | 0.657 |
| D | 5 [3-9] | 0.107 | 0.588 | 0.412 |

**Appendix Table 5.** Base scenario results, two-beta (n_obs=13, N=149, $R_1$=0.96 from day 35).

| Scenario | Median [IQR] | P(≥13) | P(takeoff) | P(no takeoff) |
|---|---|---|---|---|
| A | 1 [1-2] | 0.000 | 0.037 | 0.963 |
| B | 3.5 [1-7.25] | 0.076 | 0.425 | 0.575 |
| C | 3 [1-6] | 0.042 | 0.332 | 0.668 |
| D | 5 [3-9] | 0.116 | 0.585 | 0.415 |

**Appendix Table 6.** $R_0$ sensitivity sweep, single-beta (n_obs=13, N=149).

| $R_0$ | Scenario B | Scenario C | Scenario D |
|---|---|---|---|
| 0.96 | 0.001 | 0.000 | 0.003 |
| 1.19 | 0.007 | 0.001 | 0.011 |
| 2.12 | 0.092 | 0.047 | 0.128 |

*$R_0$ values are the three transmissibility anchors reported by Martínez et al. 2020 (post-intervention, outbreak-wide, and pre-intervention); in the single-beta model each is applied as a* *constant $R_0$ for the full simulation window. The $R_0$ = 2.12 row comes from an independent set of 1,000 simulations generated during the sensitivity sweep; the base-scenario results in Appendix Table 4 come from a separate set of 1,000 simulations at the same parameter values. Differences between the two (≤2.1 percentage points; for Scenario D, P(≥13) = 0.107 vs 0.128) reflect Monte Carlo variation at 1,000 simulations per condition, not any difference in model specification. Scenario ordering (D > B > C) is preserved in both.*

**Appendix Figure 8.** P(epidemic takeoff): single-beta vs two-beta comparison.

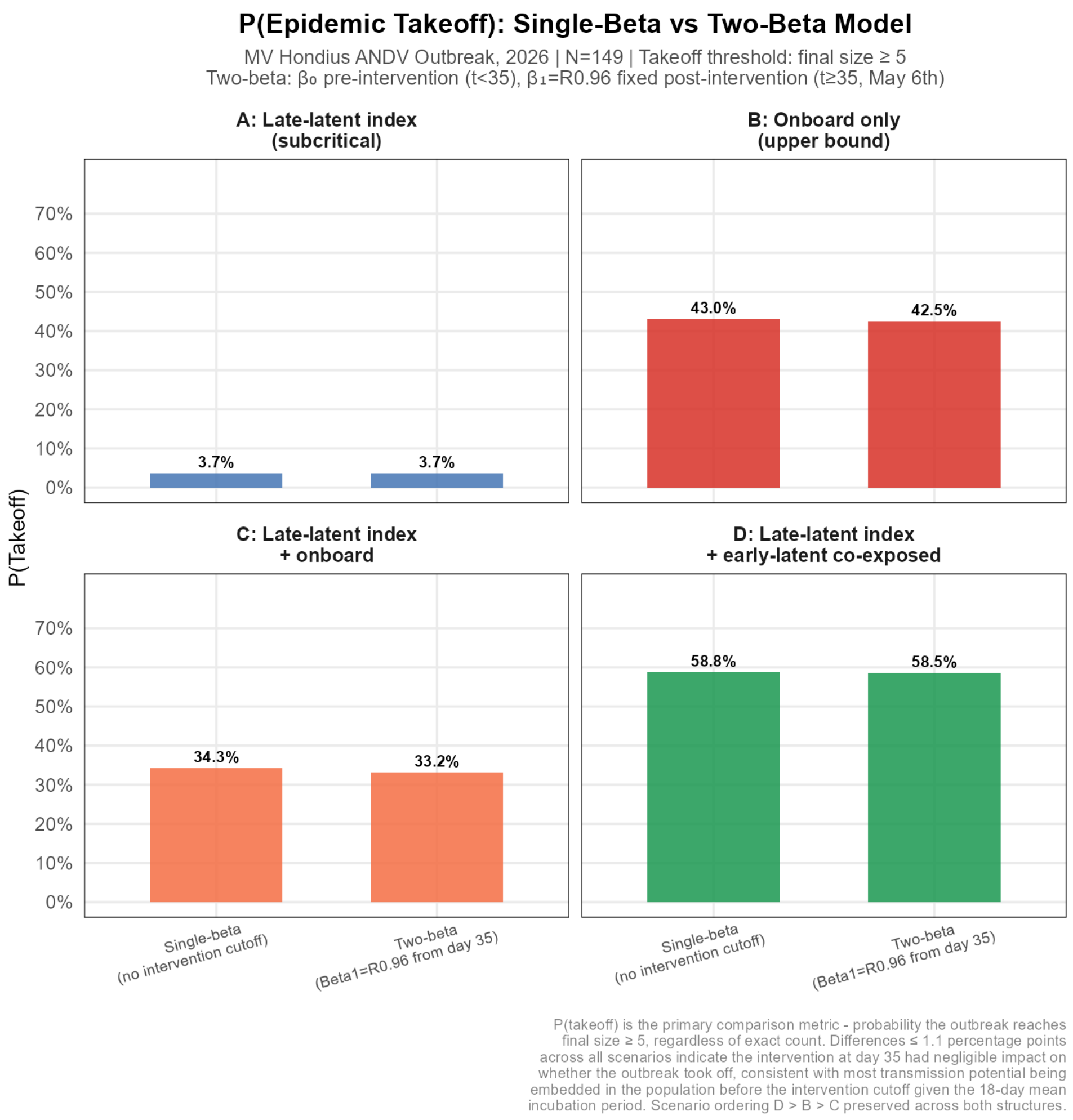

**Appendix Figure 9.** P(final size ≥13): single-beta vs two-beta comparison.

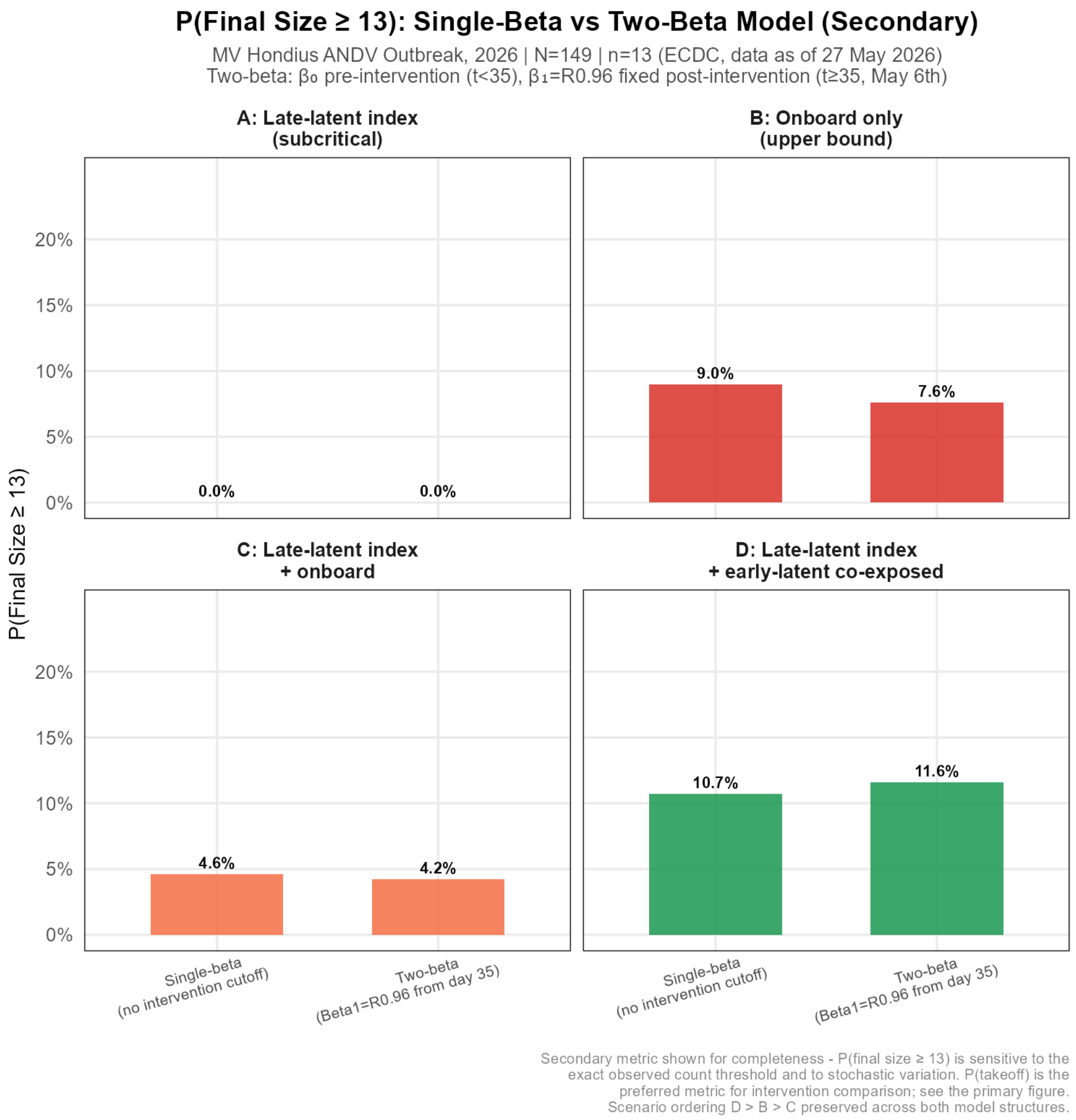

**Appendix Table 7.** Single-beta vs two-beta comparison (n_obs=13, N=149).

| Scenario | P(take) 1-beta | P(take) 2-beta | Diff | P(≥13) 1-beta | P(≥13) 2-beta | Diff |
|---|---|---|---|---|---|---|
| A | 0.037 | 0.037 | 0.000 | 0.000 | 0.000 | 0.000 |
| B | 0.430 | 0.425 | -0.005 | 0.090 | 0.076 | -0.014 |
| C | 0.343 | 0.332 | -0.011 | 0.046 | 0.042 | -0.004 |
| D | 0.588 | 0.585 | -0.003 | 0.107 | 0.116 | +0.009 |

# Appendix Results - Simulation-Based Inference (ABC)

## $R_0$ prior sensitivity analysis

Two prior upper bounds were evaluated: U(0.5, 3.0) and U(0.5, 4.0). The distance metric was RMSE between observed and simulated weekly incidence across 8 Monday-start weekly calibration bins; the composite metric (which additionally conditions on observed final size and peak) was not used as it introduces circularity. The top 1% of 100,000 simulations were retained as the approximate posterior in each run.

**Appendix Table 8.** ABC prior sensitivity (RMSE-based distance metric), two $R_0$ prior upper bounds (Epydemix, 100,000 simulations, top 1%).

| Parameter | U(0.5, 3.0) | U(0.5, 4.0) |
|---|---|---|
| $R_0$ posterior mean | 1.938 | 2.271 |

| Parameter | U(0.5, 3.0) | U(0.5, 4.0) |
|---|---|---|
| $R_0$ posterior median | 1.921 | 2.182 |
| $R_0$ 95% CrI | [0.746, 2.926] | [0.848, 3.892] |
| Boundary hit | Yes | Yes |
| P($E_1$=0) | 0.411 | 0.411 |
| P($E_1$=1) | 0.589 | 0.589 |
| P($E_3$=0) | 0.020 | 0.027 |
| P($E_3$=1) | 0.354 | 0.385 |
| P($E_3$=2) | 0.626 | 0.588 |
| P($E_1$=1, $E_3$=1) | 0.231 | 0.235 |
| P($E_1$=1, $E_3$=2) | 0.338 | 0.327 |
| Posterior mode pair | (1, 2) | (1, 2) |
| MAE (weekly incidence) | 0.500 | 0.500 |
| RMSE (weekly incidence) | 0.866 | 0.866 |
| $\text{Coverage}_{90}$ | 1.000 | 1.000 |
| Median simulated total | 10.0 | 11.0 |
| 90% PI simulated total | [7.0, 15.1] | [7.0, 16.0] |
| P(final size ≥ 13) | 0.252 | 0.275 |

*"Boundary hit" indicates the $R_0$ posterior 95% CrI upper bound approached the prior upper boundary, regardless of prior specification indicating weak $R_0$ identifiability. Initial condition posteriors were stable across both prior specifications: P($E_1$=1) was 0.589 in both runs, P($E_3$ =2) was 0.626 and 0.588 respectively, and the posterior mode pair ($E_1$=1, $E_3$=2) was invariant. The perfect coverage of 90% posterior predictive interval should be interpreted with caution. The wide intervals reflect stochastic uncertainty in small outbreak dynamics.*

## Initial condition joint posterior

**Appendix Table 9.** Joint posterior over $E_1(0) \times E_3(0)$, U(0.5, 3.0) prior (Epydemix, 100,000 simulations, top 1%, RMSE-only).

| $E_1$**(0)** | $E_3$**(0)** | **Posterior proportion** |
|---|---|---|
| 0 | 0 | 0.000 (0.0%) |
| 0 | 1 | 0.123 (12.3%) |
| 0 | 2 | 0.288 (28.8%) |
| 1 | 0 | 0.020 (2.0%) |
| 1 | 1 | 0.231 (23.1%) |
| **1** | **2** | **0.338 (33.8%)** |

*Combined P($E_1$=1) = 58.9%; P($E_3$=2) = 62.6%. Compared with the uniform prior (50% and 33.3%), both are higher; however, because the prior included the no-seed state E1(0) = 0, E3(0) = 0, which cannot generate an outbreak, the corresponding viable-seed prior probabilities are 60% and 40%. On that comparison the positive support is for E3(0) = 2, whereas support for*

*E1(0) = 1 is similar to prior expectation. Results are near-identical under U(0.5, 4.0): P($E_1$=1) = 58.9%, P($E_3$=2) = 58.8%, mode pair ($E_1$=1, $E_3$=2).*

**Appendix Figure 10.** Joint posterior heatmap $E_1(0) \times E_3(0)$, RMSE-only, U(0.5, 3.0).

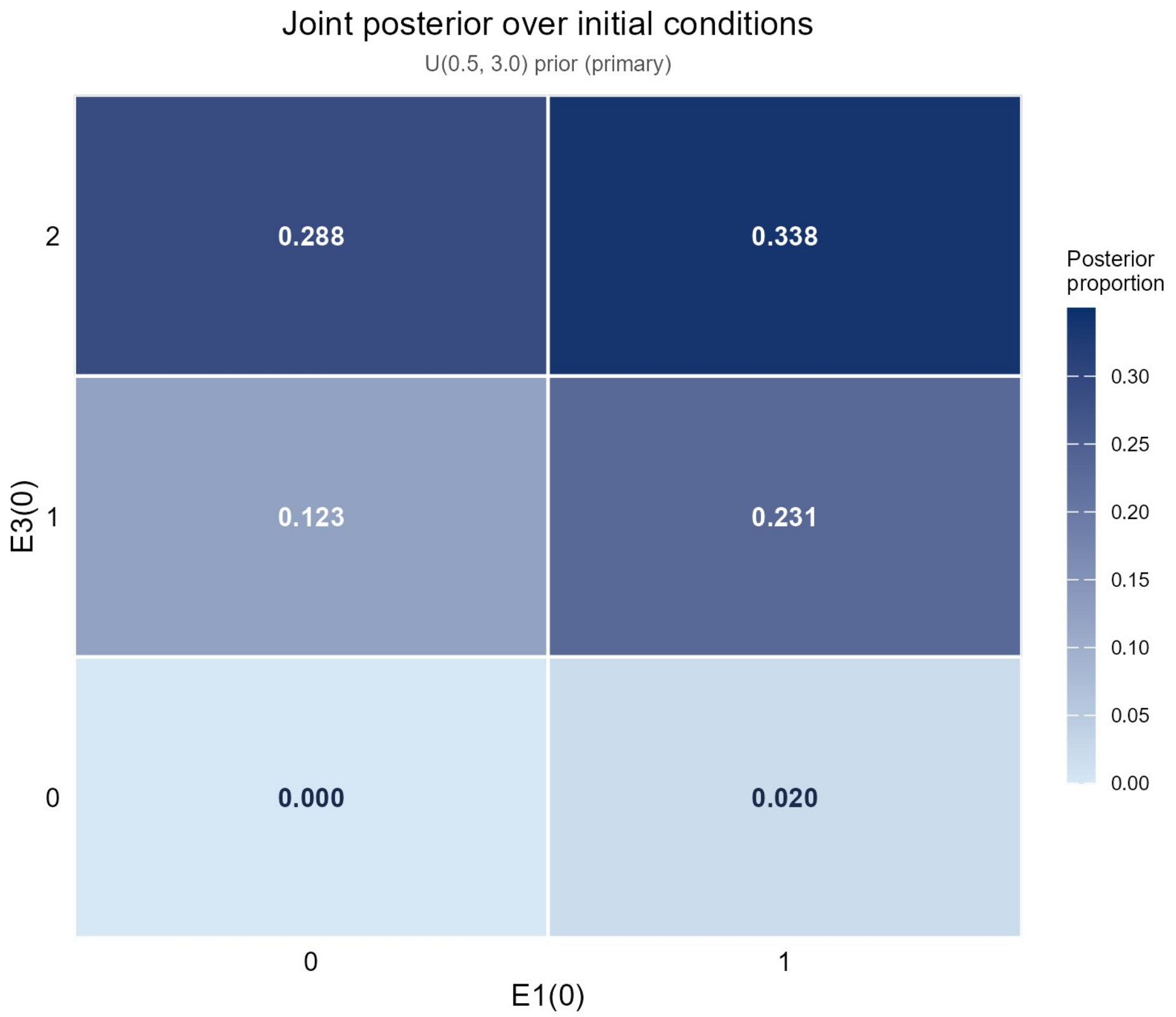

**Appendix Figure 11.** Marginal posteriors: $E_1(0)$ and $E_3(0)$

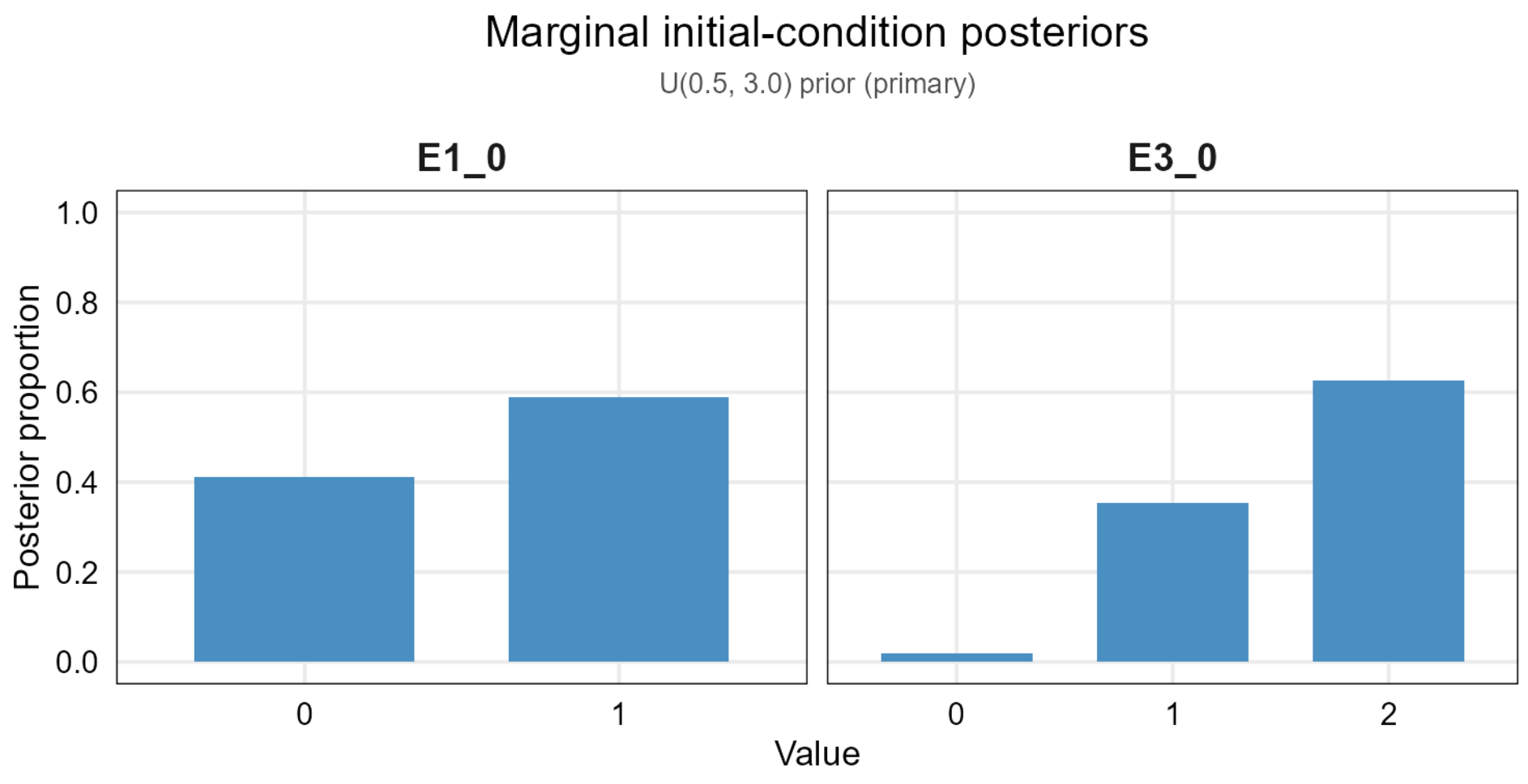

**Appendix Figure 12.** ABC posterior predictive calibration-date weekly incidence fit. Observed points correspond to symptom onset when available and report or detection date otherwise.

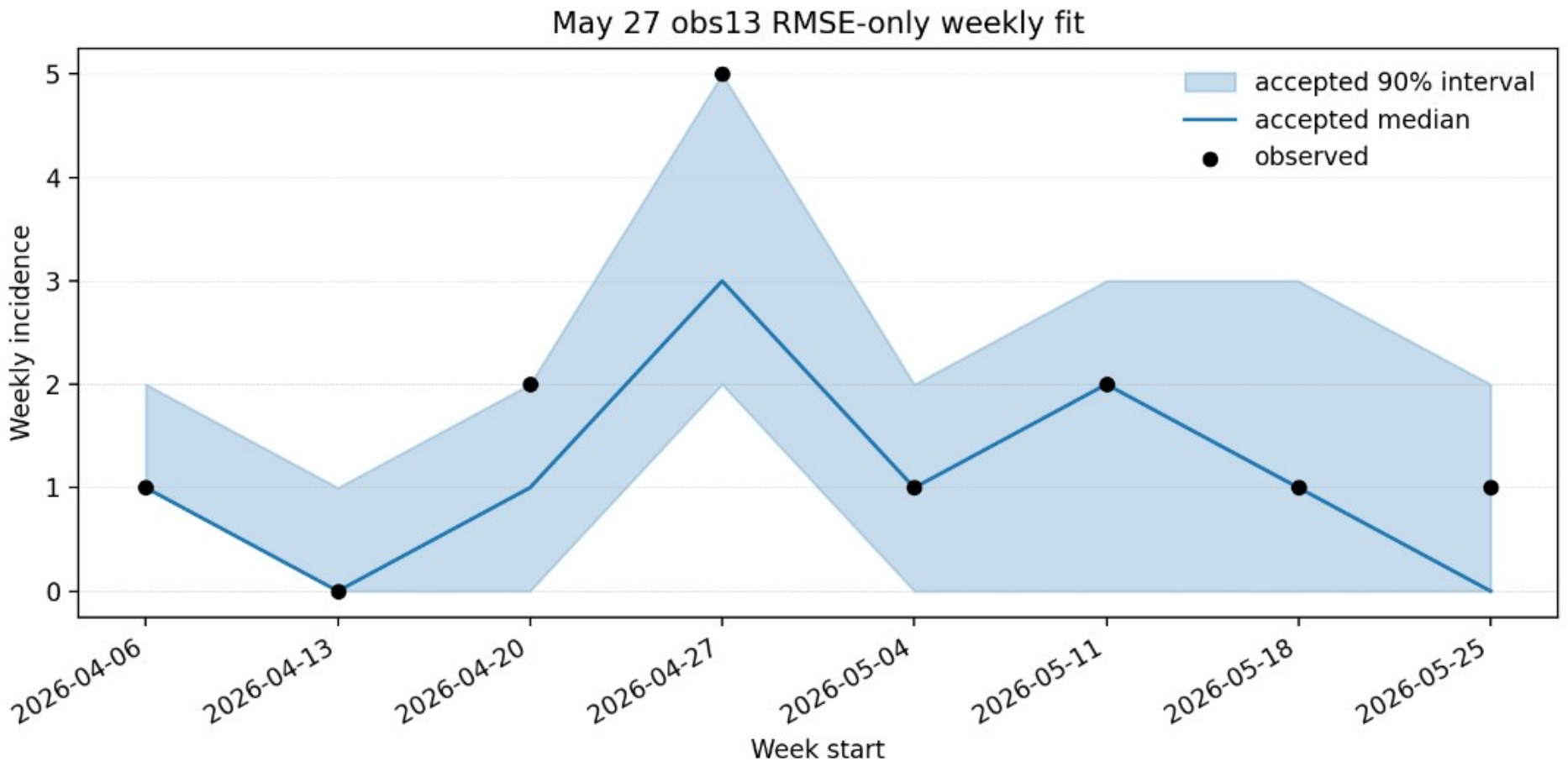

# Appendix References